\documentclass[aps,prl,superscriptaddress,floatfix]{revtex4}

\usepackage{amsmath,amssymb,url}
\usepackage{graphicx,tabularx}

\def\ie{{\it i.e.\;}}
\def\eg{{\it e.g.\;}}

\def\gev{~{\ensuremath\rm GeV}}

\def\ifb{~{\ensuremath\rm fb}^{-1}}

\def\qqquad{\qquad \qquad}

\begin{document}

\title{Understanding Single Tops using Jets}

\preprint{KA-TP-07-2009}
\preprint{SFB/CPP-09-47}

\author{Tilman Plehn}
\affiliation{Institut f\"ur Theoretische Physik, Universit\"at Heidelberg, Germany}

\author{Michael Rauch}
\affiliation{ITP, Universit\"at Karlsruhe, Germany}

\author{Michael Spannowsky}
\affiliation{ITP, Universit\"at Karlsruhe, Germany}

\begin{abstract}
  Top plus jets production at hadron collider allows us to study the
  couplings of the top quark. In the Standard Model, two single top
  processes contribute to the top-jets final state. Beyond the
  Standard Model, additional direct top production can occur. All
  three processes probe top gauge couplings including flavor
  mixing. The structure of accompanying QCD jets allows us to separate
  the direct top signal from the QCD backgrounds as well as to
  disentangle the three top plus jets production mechanisms
  orthogonally to the usual bottom tags.
\end{abstract}

\maketitle

\section{Introduction}

Experimental results from flavor physics and electroweak precision
measurements have long established the Standard Model pattern of
flavor and CP violation. The only sources of flavor and CP violation
are the Yukawa couplings, and the one essentially unknown parameter is
the relative coupling strength of the heavy third-generation quark to
the $W$ boson, \ie the CKM mixing angle
$V_{tb}$~\cite{vtb}. Its knowledge is crucial to establish
the unitarity of the CKM mixing matrix in the Standard Model.  Because
the SU(3) symmetry of QCD does not see electroweak charges, this
coupling cannot be determined in QCD-mediated top pair
production. Instead, we rely on the electroweak production process for
a single top quark in association with a quark jet to measure this
parameter of the Standard Model.\bigskip

The problem of modern particle physics is that while on the one hand
we have good reasons to expect that we will see a non-trivial
ultraviolet completion of the Standard Model at the TeV scale, we do
not know where in such an extended model this particular flavor
structure originates from.  One of the standard candidates for such
new physics at the TeV scale with all its benefits from a dark matter
candidate and stabilized Higgs mass to a valid grand unified theory is
the minimal supersymmetric Standard Model MSSM~\cite{mssm}. Mainly in
the soft-breaking terms in the MSSM Lagrangian there are multiple sources
of flavor and CP violation which naturally predict observable effects,
including flavor changing neutral currents. Assuming only one set of
supersymmetric partner states we can implement the experimental
constraints by postulating a symmetry dubbed minimal flavor
violation. This implies that there still be no sources of flavor
violation other than the Yukawa couplings, the spurions of flavor
symmetry breaking~\cite{mfv}.  An interesting
alternative might be the promotion of the gauge sector to $N=2$
supersymmetry, leading to Dirac fermion partners of the Standard Model
gauge bosons and additional scalar particles, all clearly visible at
the LHC~\cite{mrssm,sgluons}.

In this paper, we instead focus on the more subtle effects of an
approximate minimal flavor violation symmetry. Focussing on the
quark/squark sector, minimal flavor violation forces the soft squark
masses to be almost diagonal in flavor space and the scalar trilinear
$A$ terms --- for example describing squark-squark-Higgs interactions
--- to be proportional to the Yukawa couplings. Corrections consistent
with the Standard Model flavor symmetry are induced by higher powers
in the Yukawa couplings.\bigskip

Since we cannot derive minimal flavor violation from first principles,
we need to measure if and by how much it is broken. In the down-quark
sector, squarks contribute to $K$-and $B$-physics observables via
squark-gluino loops mediated by the strong coupling constant
$\alpha_s$, which gives us powerful tests of minimal flavor
violation. In contrast, in the the up-quark sector such one-loop
effects are proportional to the weak coupling $\alpha$ or the Yukawa
couplings and much harder to measure. The first-third and second third
generation mixing between the $\tilde{u}_R$ and $\tilde{c}_R$ with
$\tilde{t}_L$ squarks is essentially invisible for kaon, charm and B
experiments~\cite{flavor_orig,Dittmaier:2007uw}. Integrating out the
heavy supersymmetric particles, such loop contributions lead to
flavor-violating quark couplings to Standard Model gauge bosons, for
example a $u$-$t$-$g$ or $c$-$t$-$g$ vertex~\cite{top_fcnc}. At the
LHC processes involving valence quarks are generally more
interesting, so we will focus on the $u$-$t$-$g$ vertex, but its
second generation counter part can of course be treated the same
way.\bigskip

The search for this largely unconstrained effective coupling is linked
with the measurement of $V_{tb}$ through the relevant LHC
processes. While the effective gluon vertex leads to the direct
production of an isolated top quark, single top production is
accompanied by a quark jet. However, at the LHC we know that the
radiation of additional quark and gluon jets from the incoming quarks
is ubiquitous. Therefore, the question becomes: how can we tell apart
electroweak CKM effects in single top production (and its two
production mechanisms) and strong effects from non-minimal new physics
in direct top production, all including top quark decays as well as
realistic QCD effects and backgrounds.

\section{Direct tops and jets} 
\label{sec:dtop}

\begin{figure}[b]
 \begin{center}
   \includegraphics[width=0.8\textwidth]{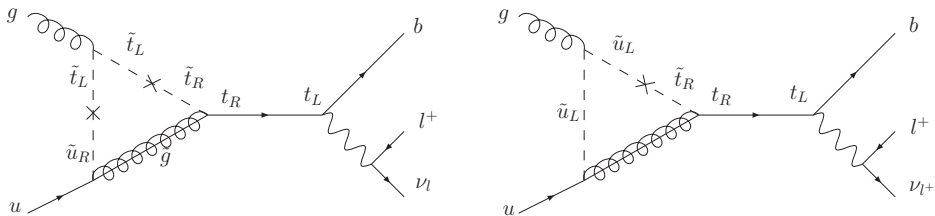} 
 \end{center}
 \caption{Sample Feynman diagrams for SUSY-induced direct top
   production. The crosses indicate left-right mass insertions which
   can mix first and third generation up squarks.}
 \label{fig:feyn_dtop}
\end{figure}

In spite of the fact that there are many flavor observables
constraining squark mixing beyond minimal flavor violation, there are
two entries in the squark mass matrices which are still largely
unconstrained. Even though we actually use the diagonalized squark
mass matrix, it is instructive to discuss the results in terms of
dimensionless mass insertions $\delta^q_{AB,ij} =
\Delta^q_{AB,ij}/\bar{m}^2$ (for the squark handedness $A,B=L,R$, the
generation indices $i,j=1...3$, and the weak isospin $q=u,d$). The
$\Delta^q_{AB,ij}$ are the relevant off-diagonal entries and
$\bar{m}^2 = m_{AA,ii} m_{BB,jj}$ the corresponding mean diagonal
entry in the squark mass matrix.\bigskip

Because we are not interested in the strongly constrained down-type
mixings $\delta^{d}$ and we focus on left-right mixing only, we denote
\begin{equation}
 \delta_{ij} \equiv \delta^u_{LR,ij}
\end{equation}
The unconstrained left-right mixing terms are the off-diagonal
$\delta_{31}$ between $\tilde{u}_R$ and $\tilde{t}_L$ and the diagonal
$\delta_{33}$~\cite{flavor_triviality,Dittmaier:2007uw}. The
left-right swapped $\delta_{13}$ instead mixes $\tilde{u}_L$ and
$\tilde{t}_R$. It is constrained by $b \rightarrow d$
transitions~\cite{rhogamma} and $\Delta m_d$ in $\bar{B}_d-B_d$
mixing~\cite{data}. The reason why the bounds on $\delta_{13}$ are
strong and those on $\delta_{31}$ hardly exist is the chargino-top
loop: if the chargino is a mix of the wino and the Higgsino, the
latter will have a large Yukawa coupling to the external bottom. The
$\tilde{u}$ instead couples to the wino content of the chargino, which
forces is to be left handed $\tilde{u}_L$. This $\delta_{13}$
constraint can only be relaxed by heavy squarks.

Assuming minimal flavor violation the generation-diagonal entry
$\Delta_{33}$ is of the general form $m_t (A_t - \mu/\tan\beta)$ and
does not need to be small. As a matter of fact, it can lead to a large
splitting of the two stop masses and ameliorates the little hierarchy
problem, so we do not expect it to be small either. It is currently
only constrained via the lower limit on the light Higgs mass and can
be measured either in stop mixing or in the minimal supersymmetric
Higgs sector~\cite{sfitter}.

The generation mixing entry $\delta_{31}$ mixes $\tilde{u}_R$ and
$\tilde{t}_L$ and at one-loop induces the flavor-changing
chromo-magnetic operator~\cite{bsm_direct_top}
\begin{equation}
m_{\tilde{g}} \; \frac{g_s}{16 \pi^2} \;
\bar{t}_{L,\alpha} \sigma_{\mu \nu} u_{R,\beta} \;
T^a_{\alpha \beta} \; G^{\mu \nu}_a  + \text{h.c.}
\end{equation}
via a squark-gluino loop. As shown on the left
Fig.~\ref{fig:feyn_dtop} this operator implies direct top production
at hadron colliders $p p \to t \to b W^+_\ell$, with a leptonic $W$
decay to avoid an undetectable purely hadronic final state.  The
corresponding partonic cross section is typically suppressed by the
heavy gluino mass in the loop and therefore proportional to
\begin{equation}
\sigma(ug \to t) \propto \frac{|\delta_{31}|^2 |\delta_{33}|^2}{m_{\tilde{g}}^2}
\end{equation}
The second diagram in Fig.\ref{fig:feyn_dtop} contributes proportional
to $|\delta_{13}|^2$, which is negligible after fulfilling all flavor
constraints. Due to the absolute values squared they are invariant
under complex phases of the squark mixing parameters. This process to
our knowledge is the only way to measure the otherwise unaccessible
$\delta_{31}$ in the era of LHC and super-$B$ factories.\bigskip

Usually, the MSSM parameter space beyond minimal flavor violation is
huge, and physical processes typically involve many possible
contributions which are free to cancel each other. In contrast, direct
top production is a strongly interacting process and depends only on
the masses $m_{\tilde{g}}$ and $m_{\tilde{t}_1}$ and the squark mixing
parameters $\delta_{33} \sim A_t$,$\delta_{31}$ and $\delta_{13}$. For
small values of $\tan\beta$ we have to include $\mu/\tan\beta$ in the
expression for $\delta_{33}$. For our numerical study we use the
SPS2~\cite{sps} inspired reference point with the GUT-scale masses
$m_0 = 1450$~GeV, $m_{1/2} =300$~GeV and $A_0=0$. The Higgs sector is
characterized by $\tan \beta=10$ and $\mu>0$. The relevant weak-scale
masses for our process are
\begin{alignat}{7}
 \tan\beta &= 9.6 \qquad & \mu &= 386 \gev \notag \\
 M_{\tilde \chi^0_1} &= 125 \gev \qquad & M_{\tilde{g}} &= 350 \gev \notag \\
 m_{\tilde{U}_{L,11}} &=m_{\tilde{U}_{L,22}} = 1538 \gev \qquad & 
 m_{\tilde{U}_{L,33}} &= 1279 \gev \notag \\
 m_{\tilde{U}_{R,11}} &= m_{\tilde{U}_{R,22}} = 1534 \gev \qquad & 
 m_{\tilde{U}_{R,33}} &= 956 \gev
\label{eq:ref_point}
\end{alignat}
%
This gives us a light (mostly) stop mass of 955~GeV and a light Higgs
mass of 117~GeV. The lightest neutralino is a viable dark matter
candidate. For the calculation of the direct top production rate at
this point we use FeynArts, FormCalc~\cite{formcalc},
LoopTools~\cite{looptools}, and HadCalc~\cite{hadcalc}.  We include
all supersymmetric QCD and electroweak contributions.\bigskip

There are four major backgrounds for direct top production $ug \to b
W^+$ with a charge identified lepton from the $W$ decay and a tagged
bottom jet with a tagging probability of 50\% and a mis-tagging
probability of 1\%
\begin{alignat}{5}
ug &\to b W^+ \qquad &&\text{irreducible and CKM suppressed} \notag \\
ug &\to d W^+ \qquad &&\text{fake bottom tag} \notag \\
\bar{d}g &\to \bar{c} W^+ \qquad &&\text{fake bottom tag} \notag \\
u \bar{d} &\to b \bar{b} W^+ \qquad &&\text{gluon splitting to two bottoms}
\end{alignat}
Of the three $W$ plus one jet backgrounds the irreducible combination
is suppressed with respect to the mis-tagging background by roughly
two orders of magnitude. Similarly, the more likely mis-tag of a charm
requires a $\bar{d}g$ initial state and should be at maximum of the
same order as the valence-quark induced $ug \to d W^+$ process. Most
importantly, the kinematic distributions of the three are very
similar, so we expect all three to vanish as a roughly constant
fraction of the leading $ug \to d W^+$ background.

The two-bottoms background process can become dangerous when the two
bottom jet are close enough to look like one bottom jet from gluon
splitting. If we require the two bottom jets to be close ($\Delta
R_{bb}<0.6$) the resulting $W^+ b\bar{b}$ rate is of the same order as
the subleading $b W^+$ production, which again means that in the
following discussion we will focus on the mis-tagged $ug \to d_b W^+$
background.\bigskip

All of the background and the signal pass the acceptance (and
triggering) cuts
\begin{alignat}{5}
 p_{T_b}  &> 20 \gev & \qqquad 
 p_{T_\ell}  &> 15 \gev \qqquad
 |\eta_b|, |\eta_\ell|  &< 2.5 & \qqquad
 \Delta R_{b\ell} &> 0.4
\label{eq:acc_cuts}
\end{alignat}
without a major reduction. Note that we apply bottom acceptance cuts
to the mis-tagged light jet in the background, so at this level there
will not be any light-flavor jets in the analysis.\bigskip

\begin{figure}[t]
 \begin{center}
   \includegraphics[width=0.35\textwidth]{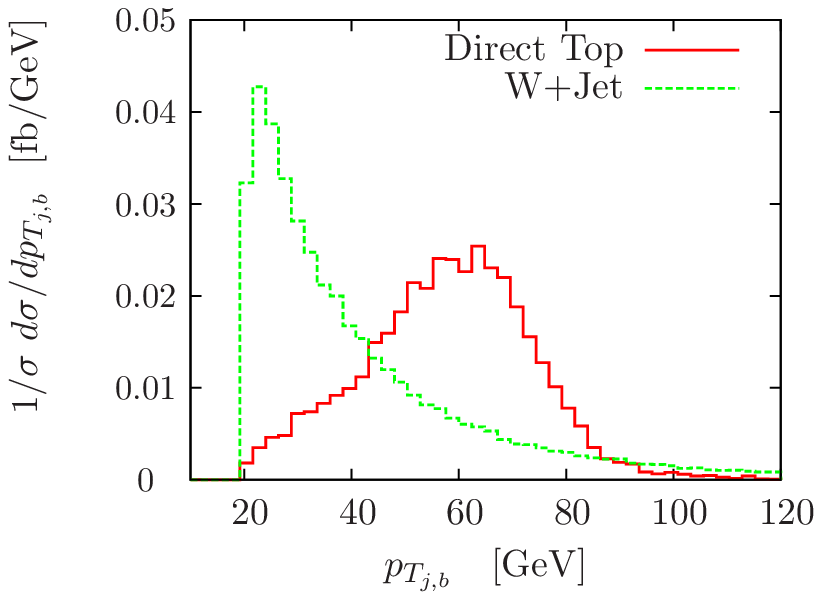} 
   \hspace*{0.15\textwidth}
   \includegraphics[width=0.35\textwidth]{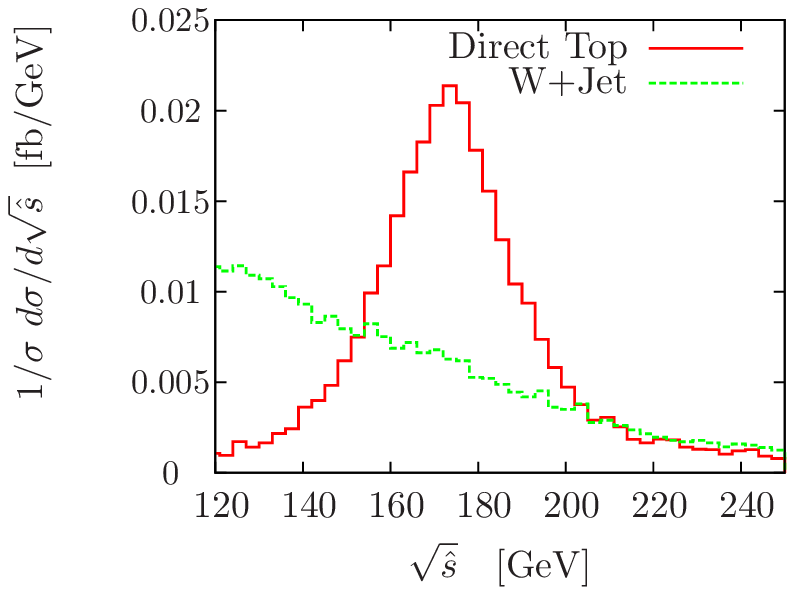} 
 \end{center}
 \caption{Normalized distributions for direct top production and the
   main background $pp \to \nu \ell^+$+jets, after acceptance
   cuts. These distributions (and only these) are generated without
   QCD jet radiation but including momentum smearing to account for
   detector effects.}
 \label{fig:dtop_rec}
\end{figure}

There are two key distributions to separate direct top production from
the continuum backgrounds, shown in Fig.~\ref{fig:dtop_rec}: due to
the signal kinematics, the transverse momentum of the bottom jet is
strongly peaked around $p_{T_b} \sim (m_t^2 - m_W^2)/(2m_t) \simeq
67.5 \gev$, even including detector effects. For the signal, the
partonic center of mass energy $\sqrt{\hat{s}}$ should also peak
around the top mass.
The unmeasured longitudinal neutrino momentum we compute from the mass
shell condition $m_{l \nu} = m_W$. The remaining two-fold ambiguity
we resolve by choosing the better top mass
reconstruction~\cite{bsm_direct_top}. These kinematic features we
exploit by requiring
\begin{alignat}{4}
  55 \gev < p_{T_b} < 80  \gev \qquad \qquad 
 165 \gev <  \sqrt{\hat{s}_\text{rec}} < 185  \gev
\label{eq:rec_cuts}
\end{alignat}
As shown in Table~\ref{tab:dtop} the statistical significance after
cutting on the kinematic features of the signal is not sufficient to
extract single top production at the LHC, even though the Gaussian
significance for $60 \ifb$ of data looks promising. Theory and other
systematic errors require a reasonable value of at least $S/B
\gtrsim 1/10$ for such a counting experiment, which means we have to
search for additional ways to extract direct top production out of the
QCD background.\bigskip

\begin{table}[b]
\begin{center}
\begin{tabular}{l|rrr|rr}
  & $\sigma_S$ & $\sigma_{Wj}$ & $\sigma_{Wb}$ & $S/B$ & $S/\sqrt{B}$ \\ \hline
after acceptance cuts & $50$~fb   & $12944$~fb & $105$~fb 
                      & 1/260   & 3.4   \\
after resonance cuts  & $13.2$~fb & $496$~fb   & $4.4$~fb 
                      & 1/38    & 4.6   \\
requiring second jet  & $7.2$~fb  & $160$~fb &
                      & 1/22    & 4.4   \\
after jets cuts       & $5.0$~fb  & $48$~fb &
                      & 1/9.6   & 5.6   \\
\end{tabular}
\end{center} \vspace*{0mm}
\caption[]{Signal and background rates for direct top production after
  acceptance cuts eq.(\ref{eq:acc_cuts}), resonance cuts
  eq.(\ref{eq:rec_cuts}), existing additional QCD jets
  eq.(\ref{eq:jets_cuts}), and finally correlation cuts on this QCD
  jet activity eq.(\ref{eq:qcd_cuts}). The statistical significances
  assume $60\ifb$ of luminosity at 14~TeV. For the merged sample we
  only consider the leading background.}
\label{tab:dtop}
\end{table}

\begin{figure}[t]
 \begin{center}
   \includegraphics[width=0.30\textwidth]{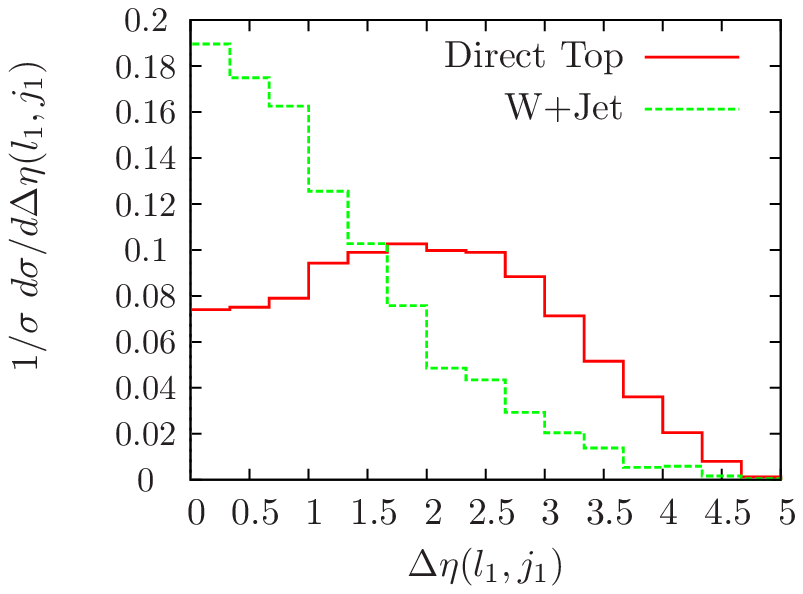} 
   \hspace*{0.02\textwidth}
   \includegraphics[width=0.30\textwidth]{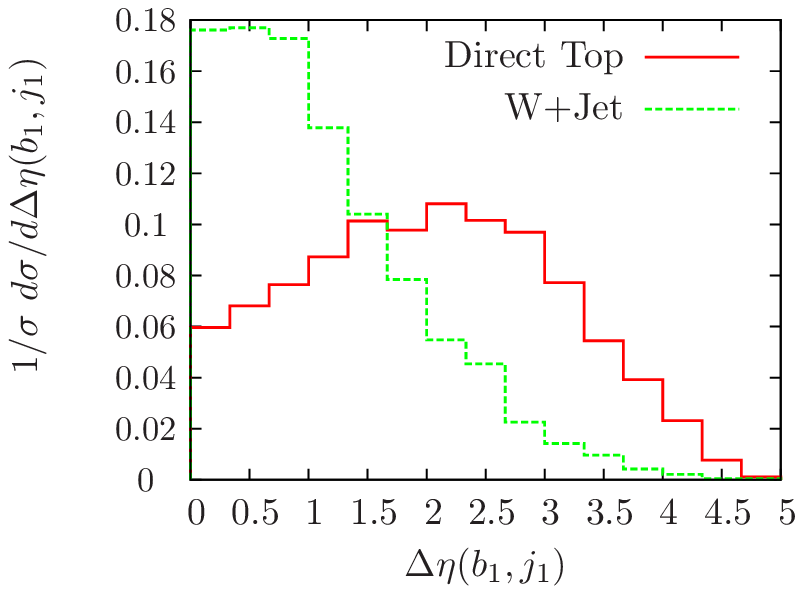} 
   \hspace*{0.02\textwidth}
   \includegraphics[width=0.30\textwidth]{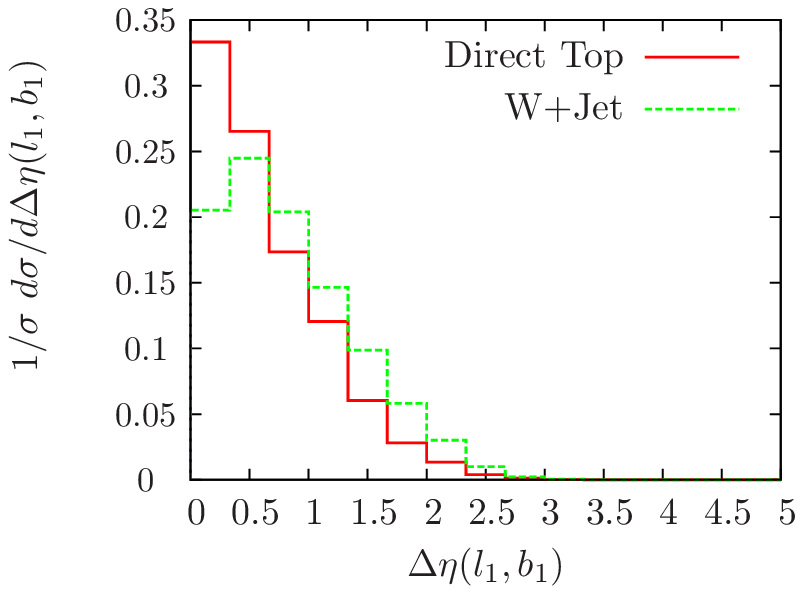} \\[3mm]
   \includegraphics[width=0.30\textwidth]{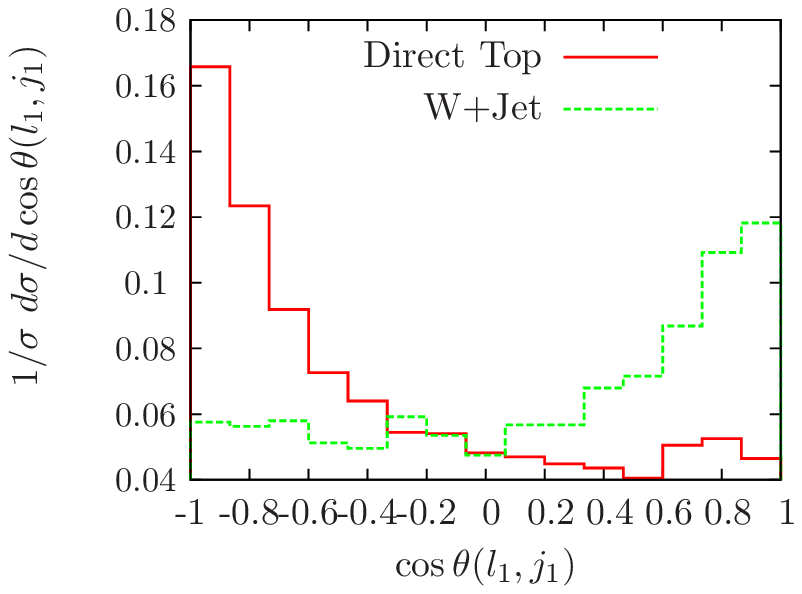} 
   \hspace*{0.02\textwidth}
   \includegraphics[width=0.30\textwidth]{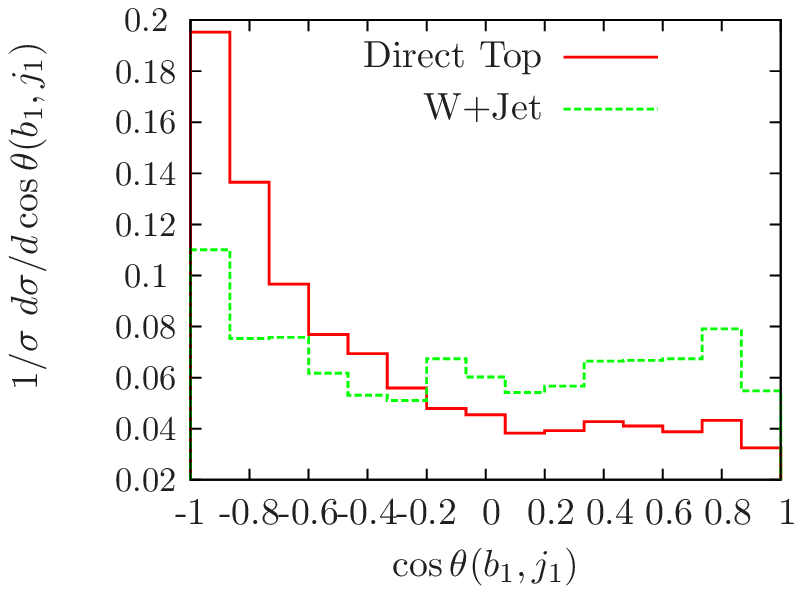}
   \hspace*{0.02\textwidth}
   \includegraphics[width=0.30\textwidth]{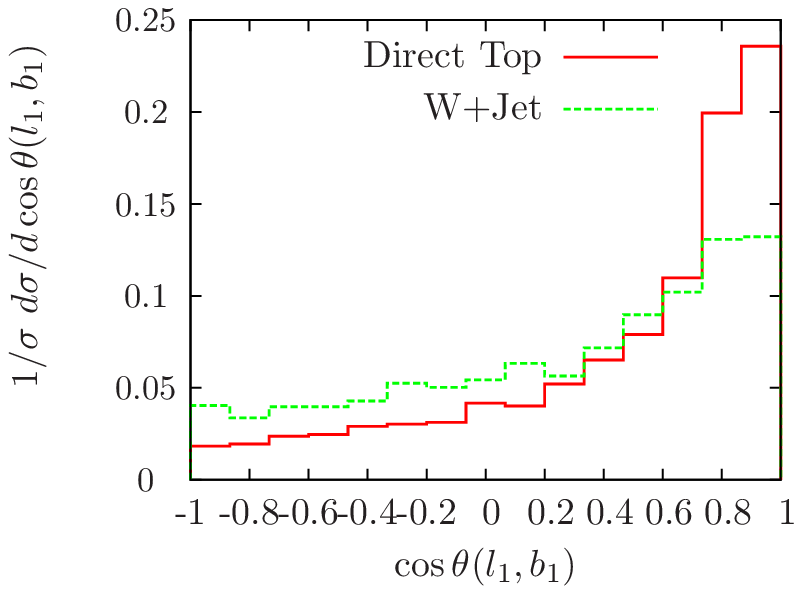}
 \end{center}
 \caption{Correlations of the first radiated QCD jet with the
   particles from the hard process for the direct top signal and the
   QCD background, after applying all acceptance and the top resonance
   cuts. We show the (similar) behavior of the pseudorapidity difference and
   the full opening angle.}
 \label{fig:dtop_j1}
\end{figure}

A distinguishing feature of direct top production which has nothing to
do with the decay kinematics is that coming from a $ug$ initial state
the produced top quark will be boosted longitudinally following the
direction of the valence quark. We can follow this behavior by
noticing that for the signal the lepton and in particular the bottom
quark are moving into the forward direction, peaked around pseudorapidities
of two. In contrast, the QCD background behaves like Drell-Yan
$W$ production with one parton splitting in the initial state, where
the $W$ boson as well as the lepton are central. Interestingly enough,
the radiated mis-tagged jet above our $p_T$ threshold is also fairly
central. The problem with this general kinematic feature is that it
is not strong enough to allow for an efficient cut in our
analysis. This changes once we include further QCD jet
radiation.\bigskip

To simulate the radiation of QCD jets beyond the leading jet we employ
the MLM matching scheme as implemented in
MadEvent~\cite{mlm,madevent}, which allows us to consistently add $pp
\to t + \text{jets}$ samples with an arbitrary number of additional
jets~\cite{ckkw}. For large transverse momenta these jets will be
correctly described by the hard matrix element and for small
transverse momenta also correctly by the parton shower. While the main
motivation for such jet merging simulations are for example $W$+jets
samples as backgrounds to Higgs and new physics searches, the same
method allows us to simulate jet radiation accompanying heavy states
at the LHC and exploit these patterns to improve the signal
extraction~\cite{mlm_heavy,sgluons}.

First, we ask for additional jet from QCD radiation which have to pass
the staggered jet acceptance cuts
\begin{alignat}{5}
 p_{T_j}  > 30,30,20 \gev  \qqquad 
 |\eta_j| < 2.5 
\label{eq:jets_cuts}
\end{alignat}
One of these three jets should be the bottom jet from the top
decay. Usually, the top decay jet will be the hardest, maybe the
second hardest jet. This means that one of the two 30~GeV jets will
have to in addition pass the top resonance cut of $p_{T_b} >55~\gev$
as listed in eq.(\ref{eq:rec_cuts}). This feature as well as our aim
to be not too dependent on the details of the jet merging simulation
motivates us to even for the light-flavor jets only consider maximum
pseudorapidities of 2.5. In this aspect, a more dedicated analysis of the
bottom tag could significantly improve the signal efficiency and hence
the Gaussian significance. While 20~GeV for the third jet sounds very
small, this analysis is also meant to show the impact the analysis of
the QCD jet activity can have on new-physics searches, so we decide to
keep it as low as possible. Depending on the structure of the measured
underlying event this threshold might have to be increased eventually.

The improvement of the direct top analysis just through requiring the
existence of two jets we show in Table.~\ref{tab:dtop}. The
higher apparent probability to radiate one additional jet from the
signal process is related to a larger number of soft jets for the
continuum background, as we will see in Fig.~\ref{fig:dtop_j2}. This
is due to the continuum structure of the background diagrams without a
hardly radiating resonant top quark and the fact that the mis-tagged
bottom is actually a massless jet and more likely to split
collinearly.\bigskip

\begin{figure}[t]
 \begin{center}
   \includegraphics[width=0.30\textwidth]{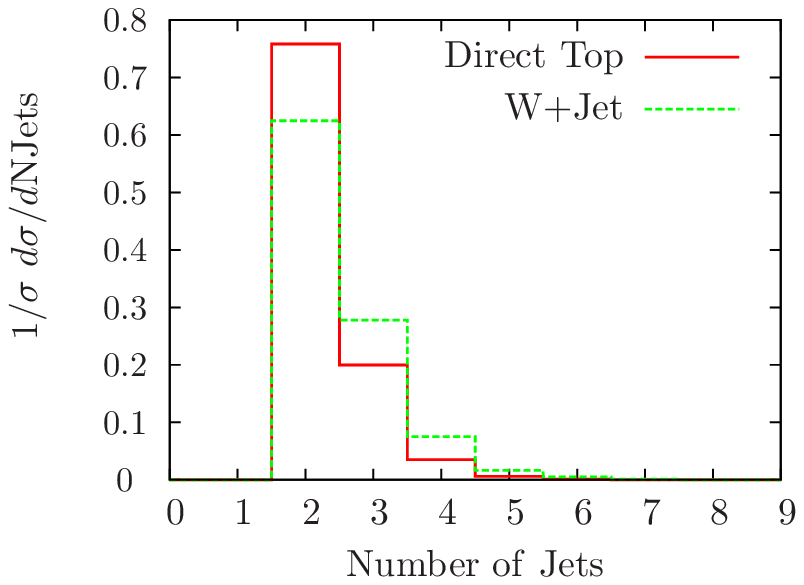} 
   \hspace*{0.02\textwidth}
   \includegraphics[width=0.30\textwidth]{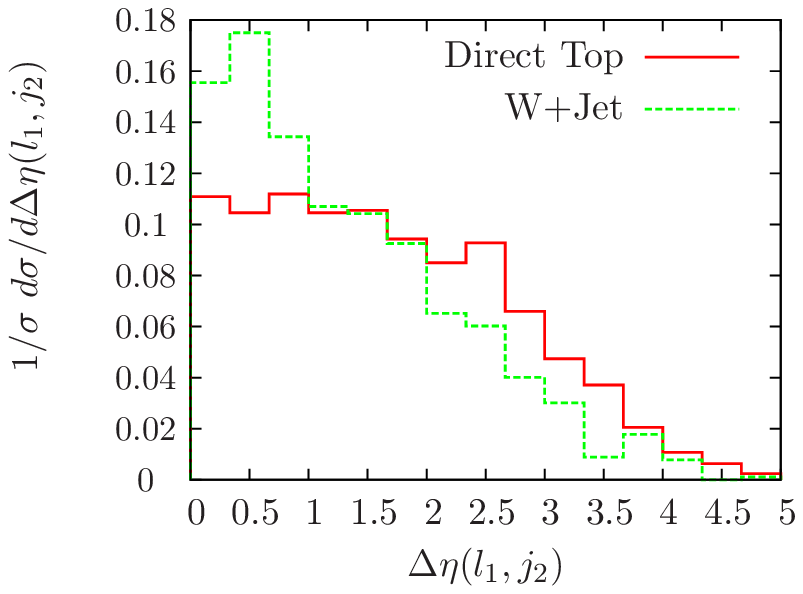} 
   \hspace*{0.02\textwidth}
   \includegraphics[width=0.30\textwidth]{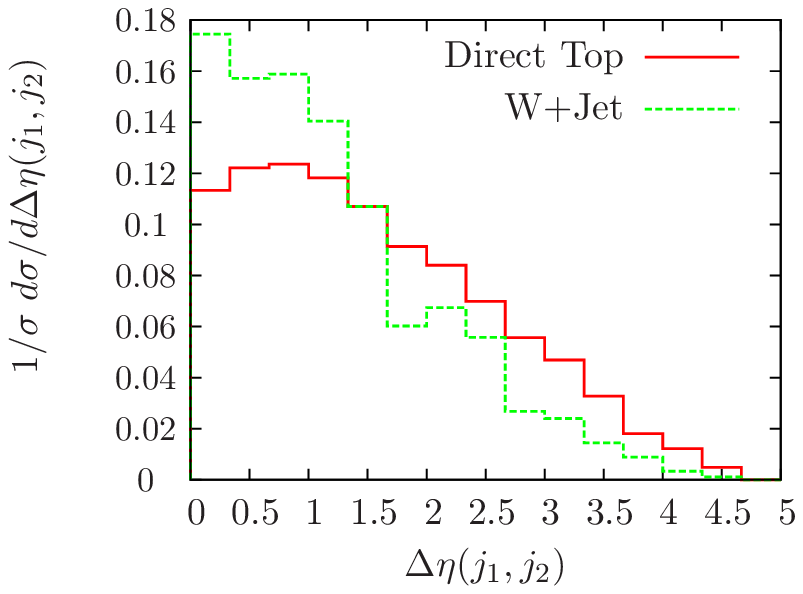}
 \end{center}
 \caption{Correlations of the second radiated QCD jet, corresponding
   to Fig.~\ref{fig:dtop_j1}.}
 \label{fig:dtop_j2}
\end{figure}

Just based on our argument above and on color factors, in the signal
case the radiation off the incoming gluon will dominate the radiation
pattern. This means that the top quark and the leading QCD jet fly
into opposite directions. We can see this behavior in the pseudorapidity
distributions as well as in the opening angle distributions in
Fig.~\ref{fig:dtop_j1}: the first radiated jet for
the signal is indeed widely separated from the lepton as well as
the bottom jet, \ie from the top quark. In contrast, the continuum QCD
background radiates jets over a wide pseudorapidity range, bounded only by
the maximum pseudorapidity of 2.5. Because the lepton is central, this means
that the pseudorapidity difference between the lepton and the first jet has
to drop off fast once we go to pseudorapidities of order two. The same is
true for the pseudorapidity difference of the first two radiated jets, once
of which is falsely tagged as a bottom jet: if both of them are
reasonably central their pseudorapidity difference is rarely going to exceed
values of 2.5.

These well distinguishable jet distributions we can now cut on, to
separate signal and background. Symmetrically, we constrain the two
pseudorapidity differences to be
\begin{alignat}{5}
 \Delta \eta_{b_1,j_1} > 1 \qqquad 
 \Delta \eta_{\ell,j_1} > 1 
\label{eq:qcd_cuts}
\end{alignat}
and show the results in Table~\ref{tab:dtop}. While these cuts do not
improve the statistical significance much beyond the top
reconstruction cuts they bring down $S/B$ to a manageable
level. Refining these jet cuts we can further improve $S/B$, but
an the expense of the number of signal events left in the
analysis.\bigskip

\begin{figure}[b]
 \begin{center}
   \includegraphics[width=0.4\textwidth]{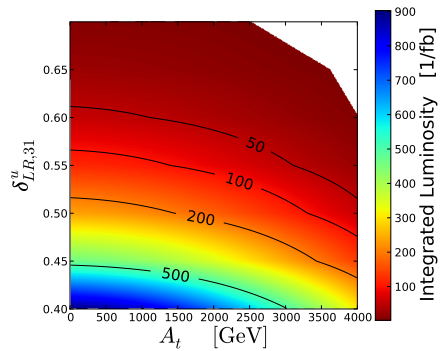} 
 \end{center}
 \caption{Necessary integrated luminosity for $95\%$ CL signal,
   assuming gaussian statistics with $S/\sqrt{B}$. The white area in
   the upper right corner is already excluded by experimental squark
   searches. We adopt a squark mass bound of $m_{\tilde{q}}>95.7
   \gev$.}
 \label{fig:dtop_coverage}
\end{figure}

Just out of curiosity we can check what happens once we include a
second QCD jet (\ie altogether three jets) in our analysis. Obviously,
this is not suitable for the full analysis, but it could give
interesting information for those signal nd background events in which
we see such an additional jet. The distribution of the number of jets
we show in the first panel of Fig.~\ref{fig:dtop_j2}.  In the jet
distributions the general pattern of the first QCD jet is still
present --- largely because it is based on the boosted nature of the
top quark in the hard signal process. In Fig.~\ref{fig:dtop_j2} we see
a similar correlation between the second QCD jet and the lepton 
as we see for the first jet. The pseudorapidity difference
between the two QCD jets is also more strongly peaked in the generally
central continuum background.\bigskip

Based on Table~\ref{tab:dtop} we can compute the $95\%$ confidence
level coverage of direct top production in the
$\delta_{31}-\delta_{33}$ plane. In Fig.~\ref{fig:dtop_coverage} we
see that with a mild dependence on $A_t \sim \delta_{33}
\sqrt{m^2_{\tilde{t},L} m^2_{\tilde{t},R}} / \left< H^u \right>$ the
LHC will be able to rule out non-minimal flavor violation through the
mass matrix entry above $\delta_{31} \gtrsim 0.5$, dependent on the
acquired luminosity.

\section{Single tops and jets} 
\label{sec:stop}

\begin{figure}[b]
 \begin{center}
   \includegraphics[width=0.9\textwidth]{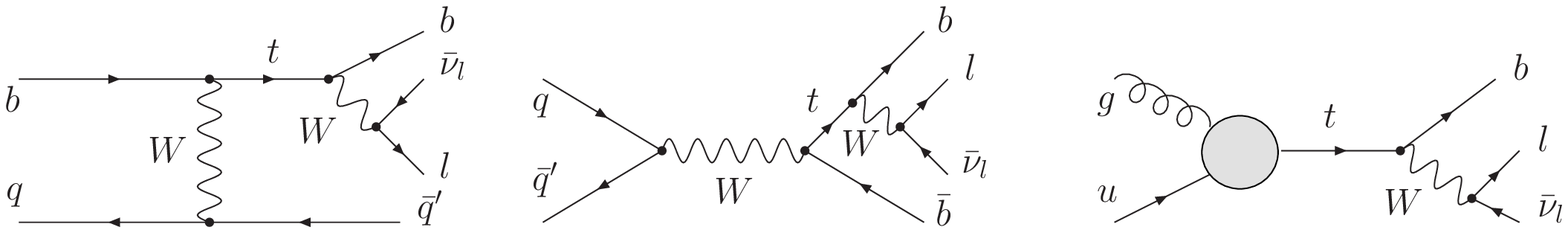} 
 \end{center}
 \caption{Sample Feynman diagrams for the three single/direct top
   production mechanisms at hadron colliders. For direct top
   production we represent the effective $ugt$ vertex with a solid
   circle.}
 \label{fig:feyn_stop}
\end{figure}

At the LHC, more signal processes contribute to the one-top-plus-jets
final state~\cite{stop_review}. The two irreducible cousins of direct
top production with jet radiation are the single top production
channels shown in Fig.~\ref{fig:feyn_stop}: one of them proceeds via a
time-like $t$-channel $W$ boson~\cite{stop_t,mcnlo} and the other
through a space-like $s$-channel $W$ boson~\cite{stop_s}. The
associated $tW$ production~\cite{stop_tw} we do not consider in this
analysis because its final state is significantly different and
neither related to QCD not irreducible from direct top production.
Encouraged by the signal vs background direct top analysis in the last
section we can ask if more generally the structure of QCD jet
radiation~\cite{stop_jets} will allow us to distinguish these three
single top and direct top production mechanisms and tell apart the
responsible coupling in or beyond the Standard Model~\cite{stop_bsm}.

Usually, the three direct/single top production channels are
distinguished using bottom tags, where aside from the top decay
products the $t$-channel process involves a forward bottom jet while
the $s$-channel process is accompanied by a central bottom jet and
direct to production does not involve additional bottom jets at
all~\cite{stop_analysis}. Already for the single top channels alone,
identifying the different QCD features first and cross checking for
bottom flavor later might allow us to improve the $V_{tb}$ measurement
in and beyond the Standard Model. Studying the forward $b$ jet in the
$t$-channel process without tagging it from the beginning also
improves our sensitivity to flavor changing $q$-$t$-$W$ couplings
enhanced by the valence quark parton
densities~\cite{stop_bsm}.\bigskip

We simulate direct top production as well as the two single top
production processes including two additional hard QCD
jets. Additional QCD jets can arise from the parton shower and are
described over their entire phase space by MLM jet merging implemented
in MadEvent~\cite{mlm,madevent}~\footnote{As the expert reader will
  observe at the end of the discussion, there is no reason why
  equivalent results could not be obtained using the MC\@@NLO
  generator for the single top channels~\cite{mcnlo}. However, direct
  top production is not available in the same framework yet.}. All
$t\bar{t}$ events appearing as part of the merged sample we
subtract. Because at this stage we are not suggesting a dedicated
analysis we first assume we know which of the jets are bottom jets and
which are light-flavor jets. For all of them we require the staggered
acceptance cuts
\begin{alignat}{5}
 p_{T_j}  > 30,30,20,20,20 \cdots  \gev  \qqquad 
 |\eta_j| < 2.5 
\label{eq:jets_stop}
\end{alignat}
The pseudorapidity range is the same for bottom and light-flavor jets and
avoids distinguishing bottoms and light-flavor jets. This way we can
later apply our analysis to all-flavor jets. To ensure that these
signal events pass the triggers we request one lepton from the top
decay with
\begin{alignat}{5}
 p_{T_\ell}  &> 15 \gev \qqquad
 |\eta_\ell|  &< 2.5 & \qqquad
 \Delta R_{j \ell} &> 0.4
\label{eq:acc_stop}
\end{alignat}
\bigskip

Without showing all the distributions we know what to expect for the
$p_T$ spectrum of the accompanying jets in the three processes: in the
$t$-channel process there will be one hard central jet balancing the
top quark and a second forward $b$ jet from the gluon
splitting. Tagging this forward bottom jet might or might not be
useful, dependent for example on the optimization of the signal vs
background analysis with respect to $S/B$ or
$S/\sqrt{B}$~\cite{bottom_densities}. This forward bottom jet will be
collinearly divergent, regularized by the bottom mass, \ie down to
transverse momenta of $p_T < 10\gev$ its $p_T$ spectrum will
diverge. The QCD jet balancing the top peaks at transverse momenta
around $40-50\gev$, almost as high as the bottom jet from the top
decay (with its Jacobian peak around 65~GeV). This we can understand
when we consider this process as one-sided weak boson fusion.

\begin{figure}[t]
 \begin{center}
   \includegraphics[width=0.30\textwidth]{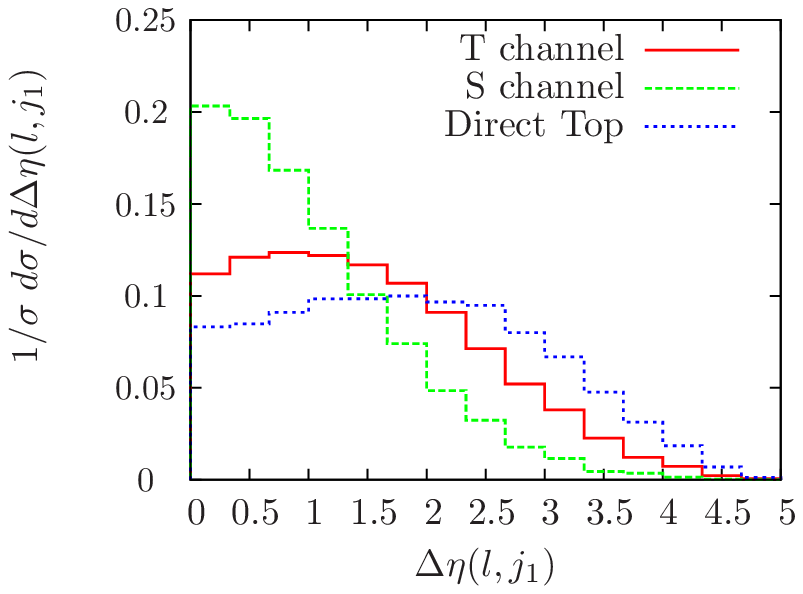}
   \hspace*{0.02\textwidth}
   \includegraphics[width=0.30\textwidth]{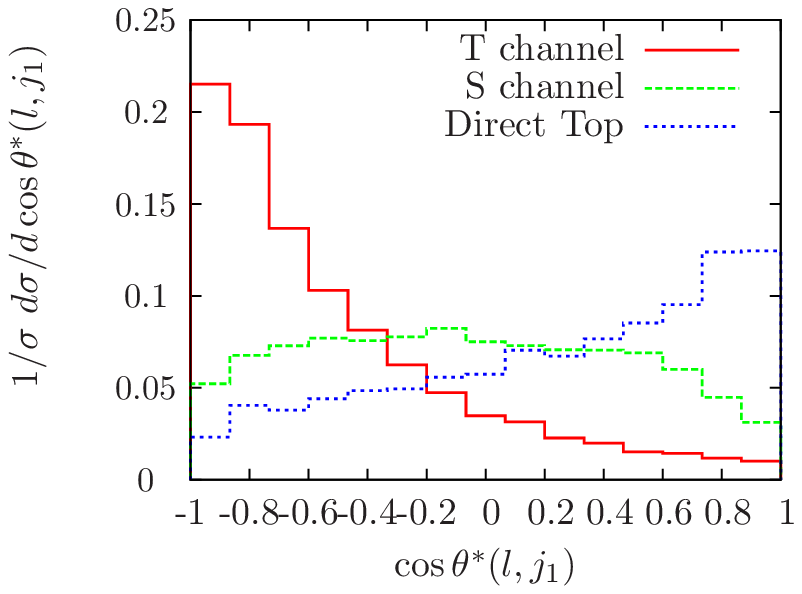}
   \hspace*{0.02\textwidth}
   \includegraphics[width=0.30\textwidth]{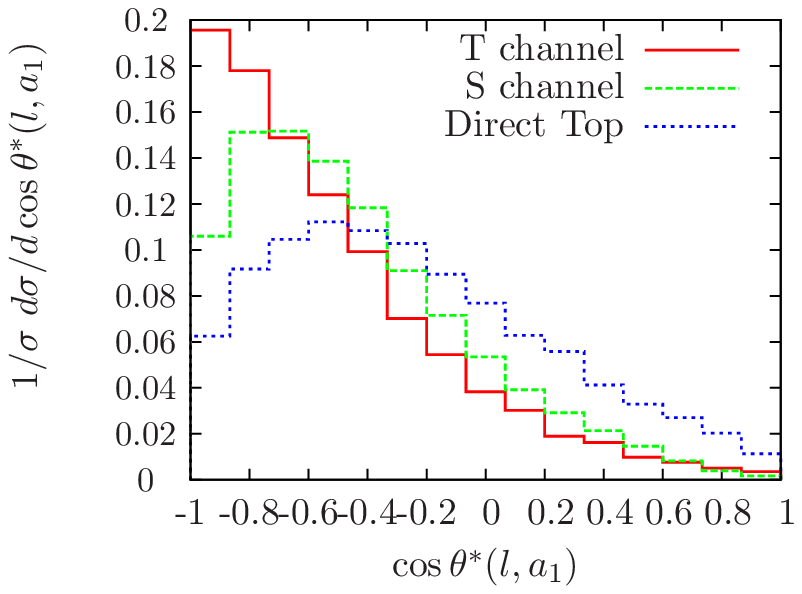} \\[3mm]
   \includegraphics[width=0.30\textwidth]{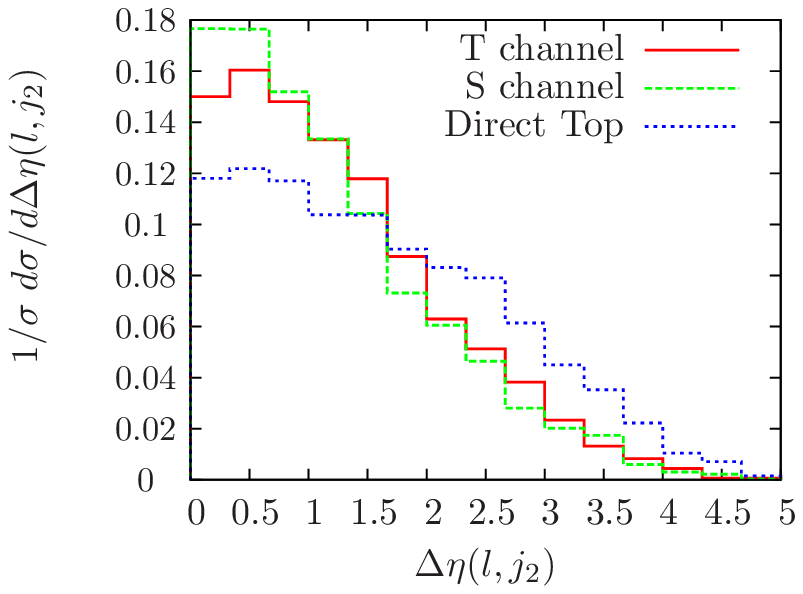} 
   \hspace*{0.02\textwidth}
   \includegraphics[width=0.30\textwidth]{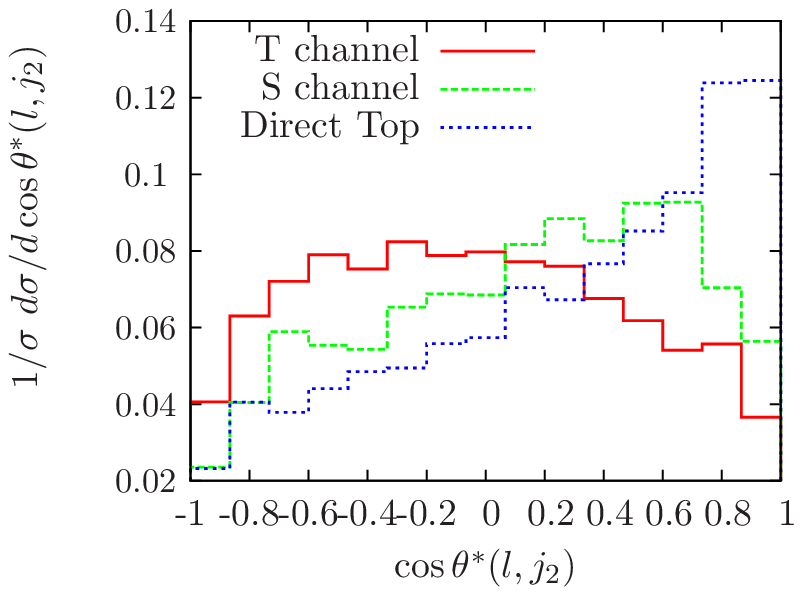}
   \hspace*{0.02\textwidth}
   \includegraphics[width=0.30\textwidth]{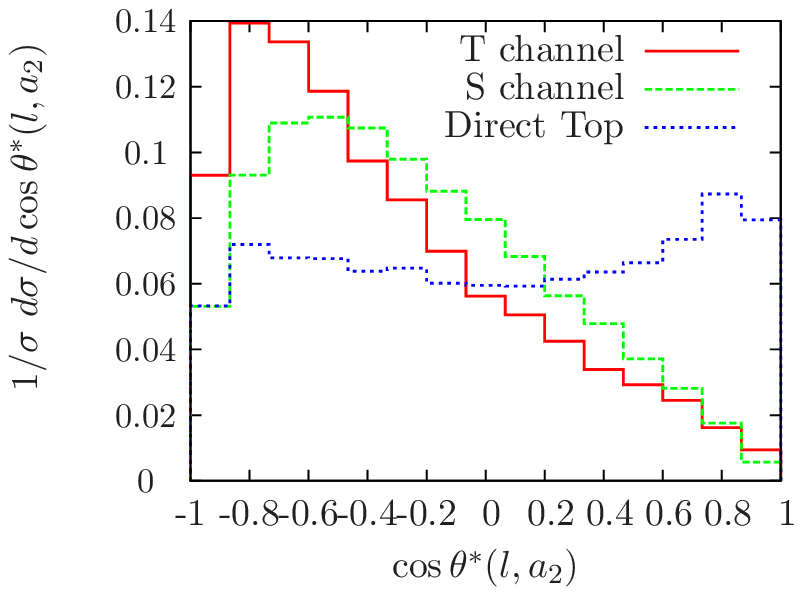} \\[3mm]
   \includegraphics[width=0.30\textwidth]{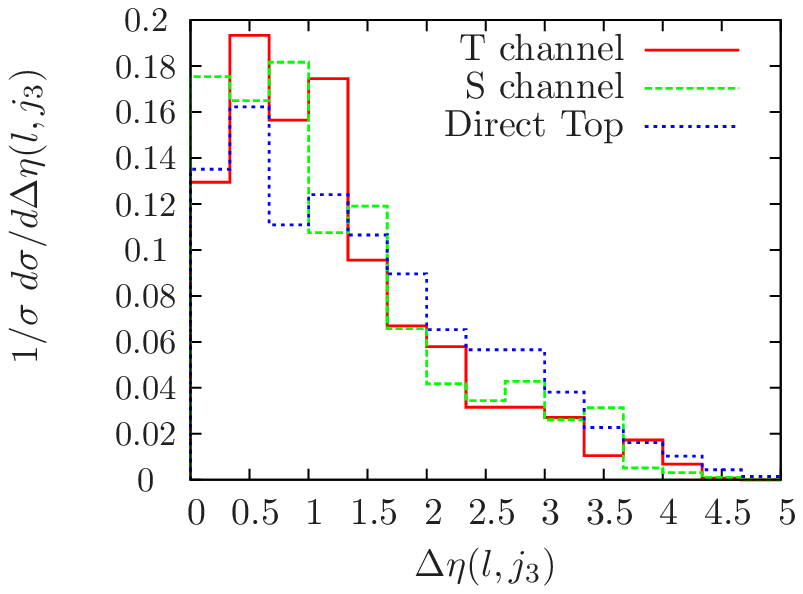}
   \hspace*{0.02\textwidth}
   \includegraphics[width=0.30\textwidth]{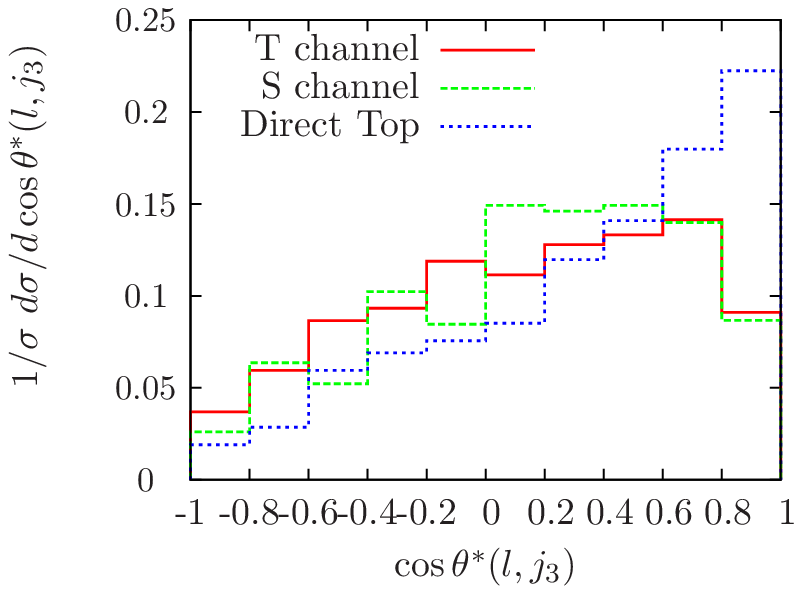} 
   \hspace*{0.02\textwidth}
   \includegraphics[width=0.30\textwidth]{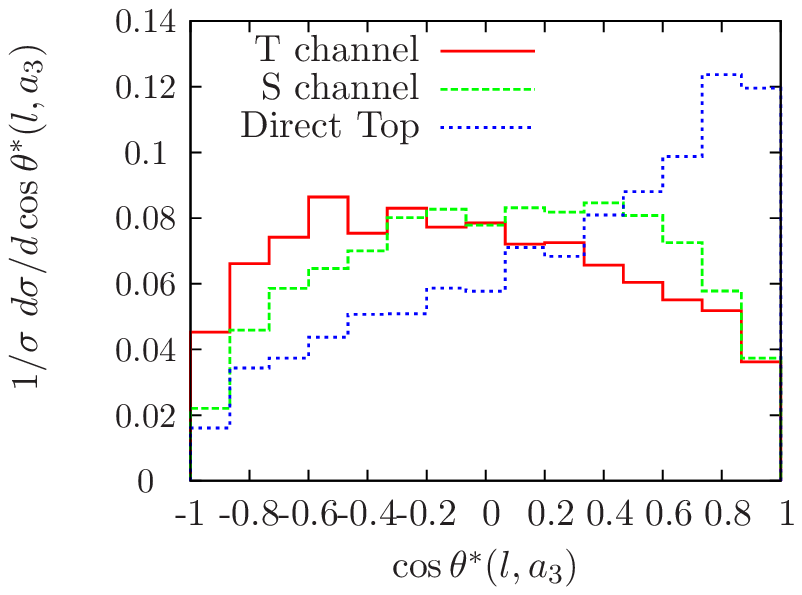} \\[3mm]
   \hspace*{0.66\textwidth}
   \includegraphics[width=0.30\textwidth]{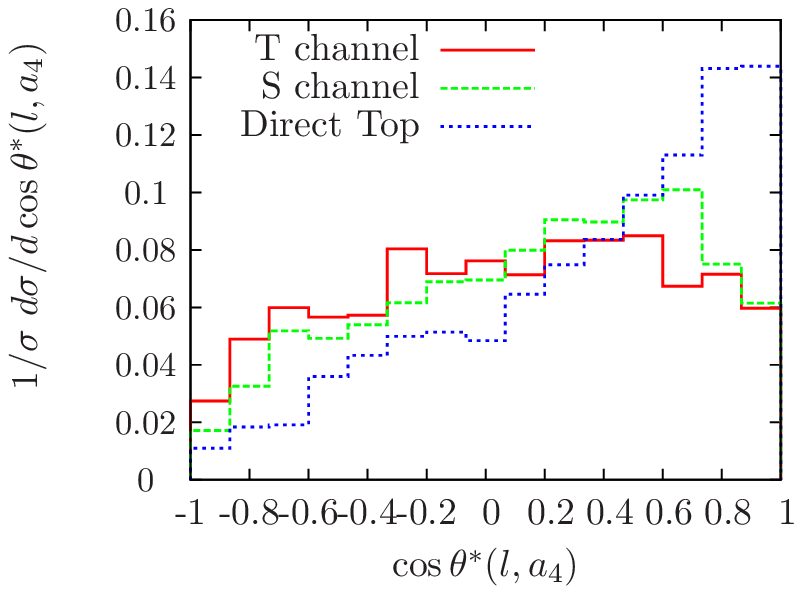} \\[3mm]
 \end{center}
 \caption{QCD jet--lepton correlations for single and direct top
   production.  We show the pseudorapidity difference as well as the $\cos
   \theta^*$ dependence as discussed in the text. The label $a$
   represents all-flavor jets which might be bottom jets ($b$) or
   light-flavor jets ($j$).}
 \label{fig:stop_j1}
\end{figure}

In the $s$ channel the flavor structure of the Standard Model enforces
a second bottom jet from the off-shell $W$ decay. This jet plays a
similar balancing role as the light-flavor jet in the $t$-channel
process. We now we see two competing hard bottom jets of comparable
$p_T$, both peaking around $40-80\gev$.

Direct top production in contrast does not predict any additional hard
jets in the detector. Because it involves a flavor changing production
vertex there should be no additional bottom jets, and the light-flavor
jets will follow the typical initial state radiation pattern.\bigskip

To describe the angular correlations of the final
state~\cite{stop_angular} including QCD jets in more detail we
consider the observable $\cos \theta^*(P_1,P_2)$. It parameterizes the
angle between $\vec{p}_1$ in the rest frame of $P_1 + P_2$ and
$(\vec{p}_1+\vec{p}_2)$ in the lab frame. It is not symmetric in its
arguments; if the two particles are back to back and $|\vec{p}_1| >
|\vec{p}_2|$ it approaches $\cos \theta^* =1$, whereas for
$|\vec{p}_1| < |\vec{p}_2|$ it becomes $-1$. In between it vanishes in
the case where $\vec{p}_1$ in the center of mass frame is orthogonal
to the lab frame movement of this center of mass.\bigskip

This behavior we confirm in the first row of Fig.~\ref{fig:stop_j1}.
For $t$-channel single top production the hardest light-flavor jet
balances the top, so the lepton is most likely back to back with the
hardest jet. Because it balances the heavy top, the leading jet is
harder than the lepton and the $\cos \theta^*_{\ell j_1}$ distribution
peaks at $-1$.  For the $s$ channel the decay lepton will be central,
as will be the first QCD jet. Except for the azimuthal angle we expect
no back-to-back configuration, which gives us $\cos \theta^*$ values
flat around zero.  In direct top production the first jet is radiated
at high rapidity. Because of the color factor it most likely comes
from the gluon, which means it will be collinear as well as soft. The
top is boosted against the incoming gluon and with it the decay
lepton. This means that the lepton and the first QCD jet are back to
back, and the hard decay lepton will be more energetic than the soft
collinear jet. The $\cos \theta^*$ distribution will then peak around
$+1$.

Considering further jets in $t$-channel single top production the
lower energy of subsequent jet radiation washes out the $\cos
\theta^*$ behavior. This is due to the role of the energy hierarchy
with respect to the lepton in this observable. Once we arrive at jet
number three all we see is a general parton shower jet radiation,
slightly forward but uncorrelated with the top decay products. For the
direct top channel the pattern of the second jet resembles the first,
because the radiation off the incoming gluon will still
dominate. Similarly, additional jets accompanying $s$-channel top
production will become slightly more forward and hence more likely to
move towards larger $|\cos \theta^*|$ values, but without an
exploitable structure.\bigskip

\begin{figure}[t]
 \begin{center}
   \includegraphics[width=0.30\textwidth]{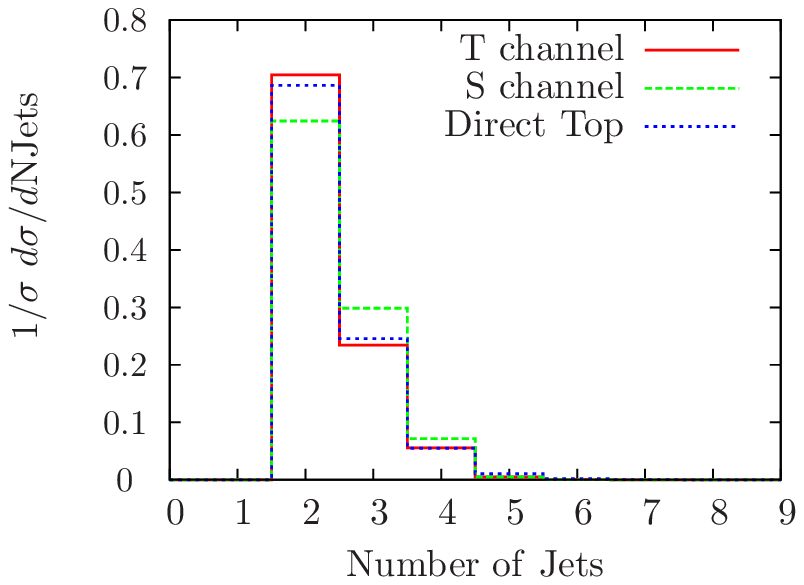}
   \hspace*{0.02\textwidth}
   \includegraphics[width=0.30\textwidth]{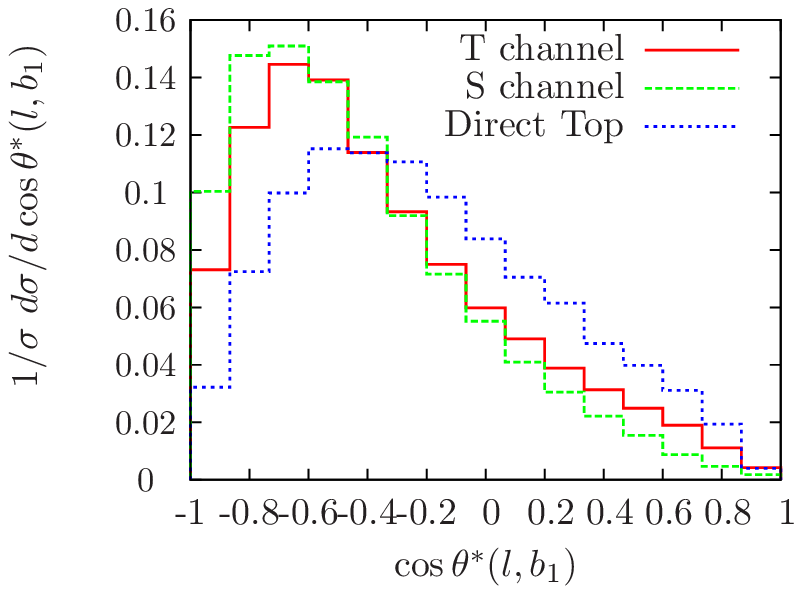}
   \hspace*{0.02\textwidth}
   \includegraphics[width=0.30\textwidth]{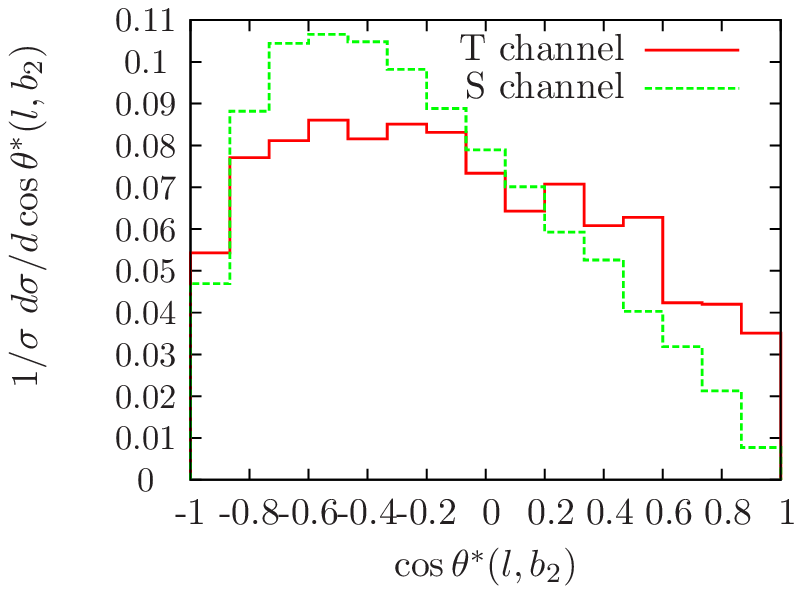} \\[3mm]
   \hspace*{0.34\textwidth}
   \includegraphics[width=0.30\textwidth]{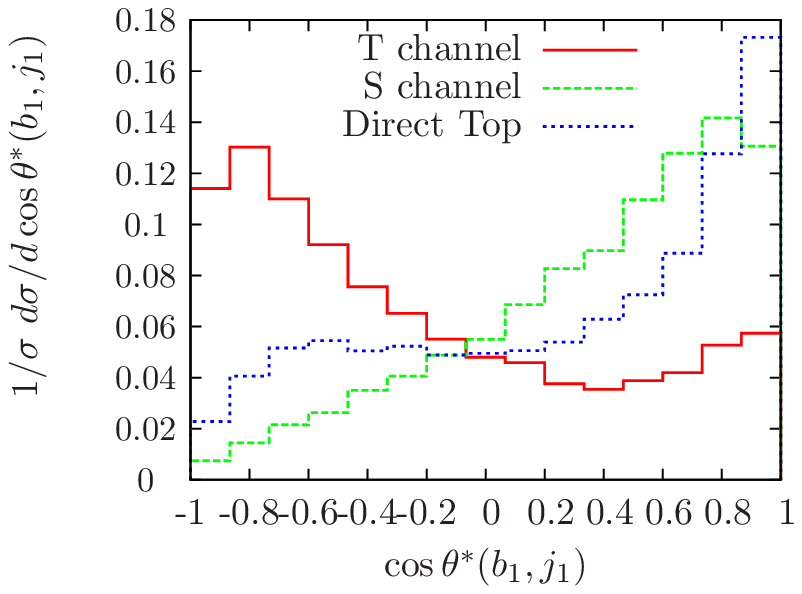}
   \hspace*{0.02\textwidth}
   \includegraphics[width=0.30\textwidth]{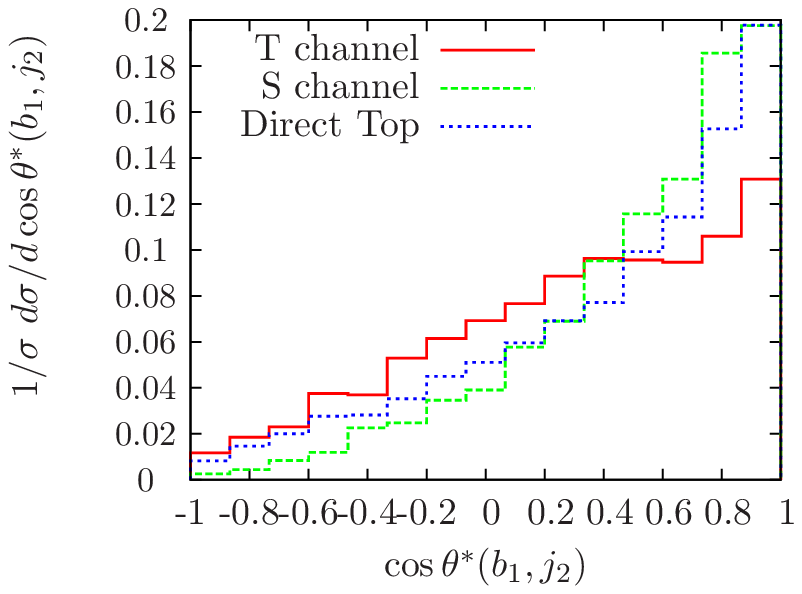} \\
 \end{center}
 \caption{Correlations of the up to two bottom jets with the lepton
   from the top decay, corresponding to Fig.~\ref{fig:stop_j1}. We
   also show the correlation of the leading bottom jet with the first
   two light-flavor jets.}
 \label{fig:stop_b1}
\end{figure}

Following our original motivation, what is most interesting are
angular correlation of light-flavor and bottom jets without assuming a
$b$ tag. In the right column of Fig~\ref{fig:stop_j1} we see that for
$t$-channel single top production the hardest all-flavor jet indeed
balances the top quark, \ie it is not bottom flavored.  The second
hardest jet then comes from the top decay, which we can check by
comparing it features with the bottom correlations shown in
Fig.~\ref{fig:stop_b1}. The third and forth jets both arise from
parton splitting and have not distinct correlations with the top decay
products --- being bottom flavored or not. Checking their $p_T$
distribution we can confirm that they simultaneously drop off fast at
values above 40~GeV.

For $s$-channel single top production both leading all-flavor jets
have little to do with the $\cos \theta^*_{\ell j_1}$ distribution,
which means they are the two bottom jets from the hard process. Again,
we confirm this behavior in Fig.~\ref{fig:stop_b1}. In this figure we
also see that essentially all bottom jets prefer values $\cos
\theta^*_{\ell b} \to -1$, which means the bottom jets balance the
lepton direction and the bottom energy lies above the lepton
energies. This is simply an effect of the intermediate $W$ decay step
which softens the $W$ decay product as compared to the bottom jet.

For direct top production the hardest jet is usually the bottom decay
jet. In addition we expect no more bottom jets, so the distributions
of $j_n$ match those of $a_{n+1}$. Because the argument of the color
factor preferring radiation off the incoming gluon combined with the
boosted center of mass frame holds in general, subsequent jets show a
similar preference for $\cos \theta^* \to 1$.\bigskip

Beyond the individual jet-lepton correlations we can also study the
correlations between the different all-flavor jets. In
Fig.~\ref{fig:stop_2d} we should sample distributions of this kind for
the three production processes under consideration.

For the $t$ channel process we already know that the two leading jets,
one light-flavor and one the bottom decay jet, both reside at small
values of $\cos \theta^*_{\ell a}$. The third and fourth jet, again
one bottom and one light-flavor, do not have a strong preference for
large values of $|\cos \theta^*_{\ell a}|$, but in contrast to the
leading two jets they show a correlation with each other, based on the
general radiation pattern.

In the $s$-channel process the two leading jets are the two bottom
jets. They are correlated and prefer small values of $\cos
\theta^*_{\ell a_j}$, dependent mostly on the lepton energy. For the
next two jets the correlations is considerably weaker, and from
Fig.~\ref{fig:stop_j1} we already know not to expect much in formation
in their $\cos \theta^*_{\ell a}$ distributions.

Direct top production produces a hard bottom jet with a slight
preference towards $\cos \theta^* \to -1$. Aside from that, the
universal jet radiation structure shows a diagonal correlation with
the slightly favored region $\cos \theta^* \to 1$, becoming much more
prominent for the two subleading jets.

\begin{figure}[t]
 \begin{center}
   \includegraphics[width=0.3\textwidth]{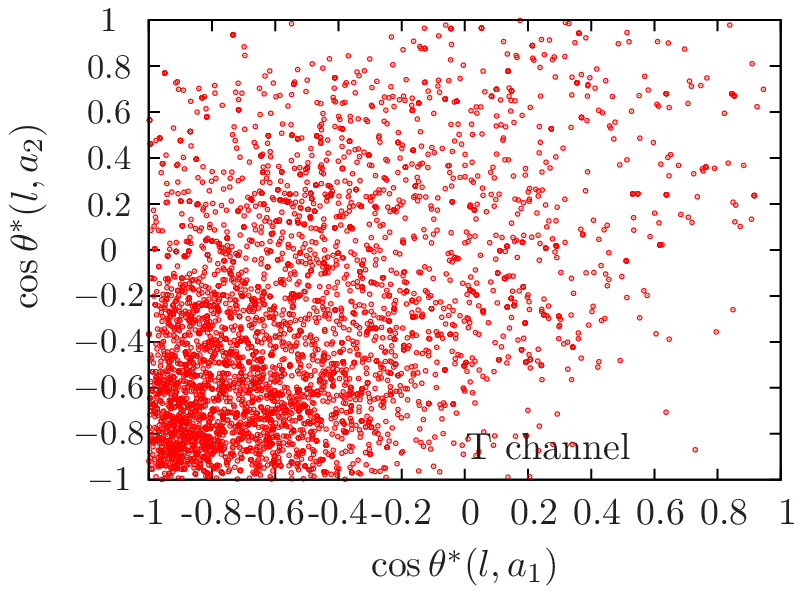} 
   \hspace*{0.02\textwidth}
   \includegraphics[width=0.3\textwidth]{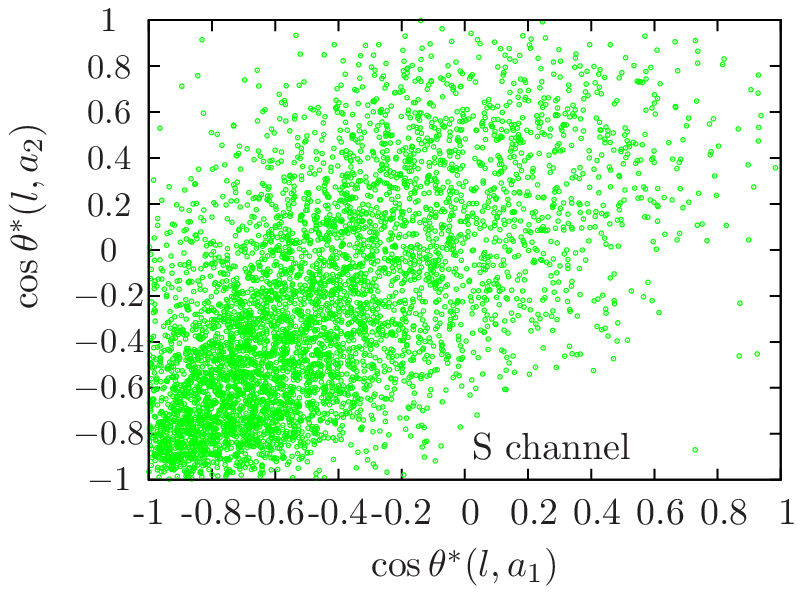} 
   \hspace*{0.02\textwidth}
   \includegraphics[width=0.3\textwidth]{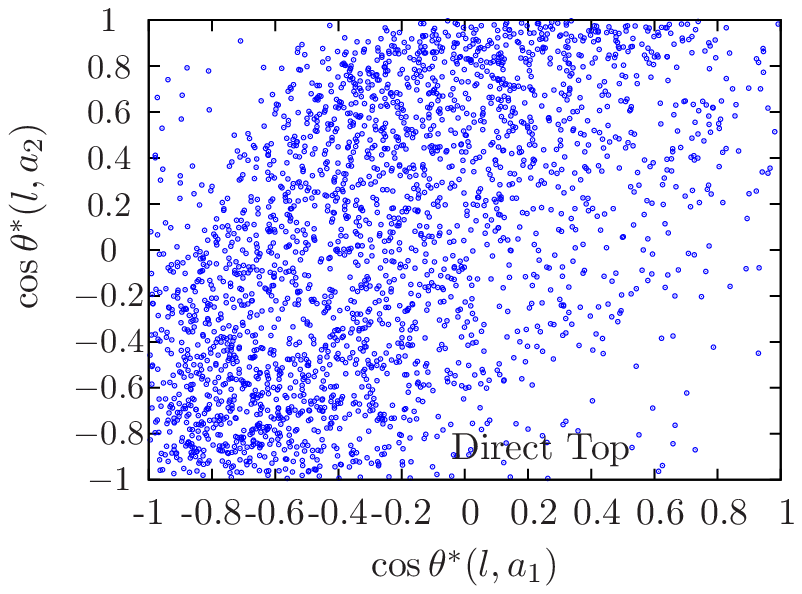}\\[3mm]
   \includegraphics[width=0.3\textwidth]{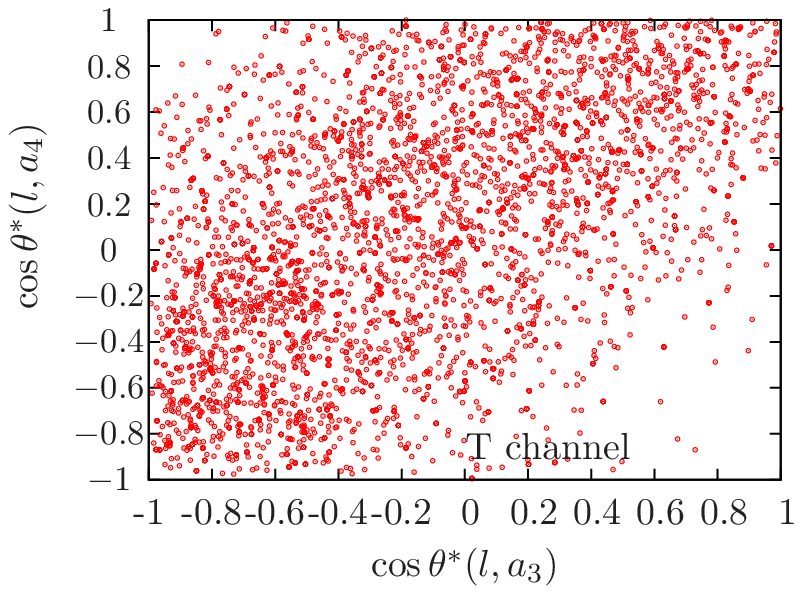} 
   \hspace*{0.02\textwidth}
   \includegraphics[width=0.3\textwidth]{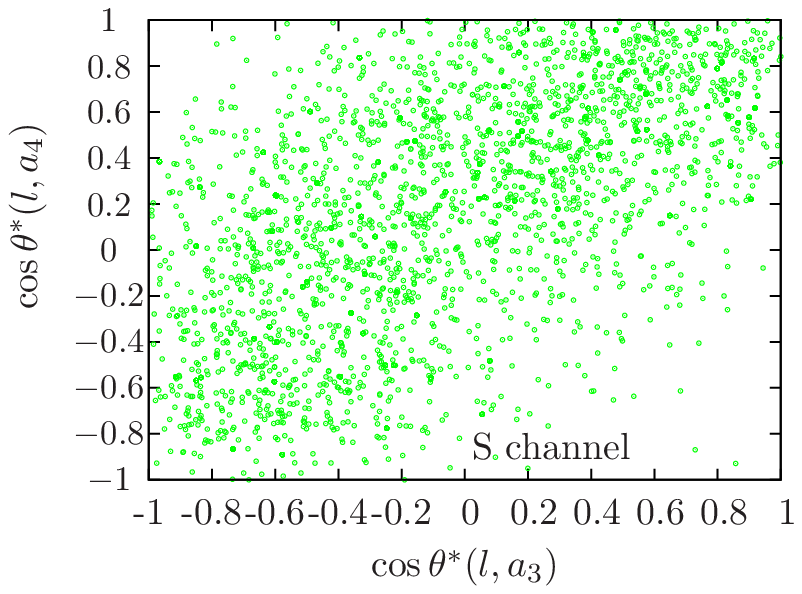}
   \hspace*{0.02\textwidth}
   \includegraphics[width=0.3\textwidth]{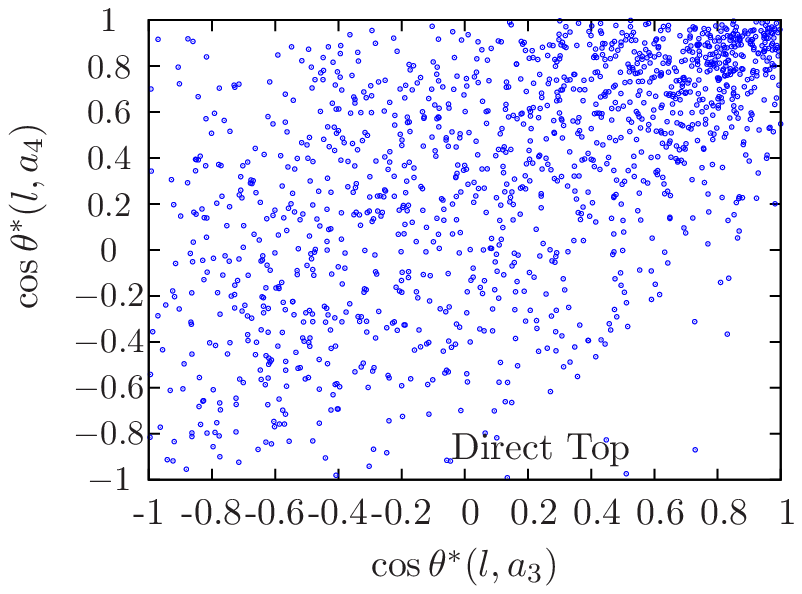}
 \end{center}
 \caption{Sample angular correlations between the $\cos \theta^*_{\ell
     a}$ for all-flavor jets. From left to right we show $t$-channel
   single top production, $s$-channel single top production, and
   direct top production.}
 \label{fig:stop_2d}
\end{figure}

\section{Outlook \label{sec:outlook}}

In this paper we have shown how we can exploit the QCD jet activity to
extract direct top production from the QCD background and to
tell apart $t$-channel single top production, $s$-channel single top
production, and direct top production.\bigskip

Direct top production is the only way to observe or constrain the
supersymmetric flavor-violating squark mass entry $\delta^u_{LR,31}$
in the near future. It is mediated by the flavor changing neutral
current $u$-$t$-$g$ interaction and has to be extracted from $W$+jets
backgrounds with a light-flavor jet mis-tagged as a bottom. While
kinematic cuts based on the top resonance structure of the matrix
element are not sufficient to produce a promising signal-to-background
ratio $S/B$, we find that additional cuts on the jet correlation from
QCD radiation improve this ratio enough to allow for a meaningful LHC
analysis.\bigskip

The same kind of QCD jet correlations with the top decay lepton allow
us to distinguish (normalized) samples of $t$-channel single top
production, $s$-channel single top production, and direct top
production. All three channels are irreducible if we consider searches
for one top quark plus jets. Our distinction is purely based on
angular correlations and serves as an orthogonal test to possible
bottom tags. In single top searches applying explicit bottom tags only
later in the analysis allows us to make use of the parton density
enhancement when looking for flavor-changing top couplings. Both
methods combined should be able to unambiguously determine the third
generation flavor structure of the Standard Model, including the CKM
mixing element $V_{tb}$ as well as new-physics effects.

\subsection*{Acknowledgments}

We thank Christoph Englert, Stefan Gieseke, Christoph Hackstein for
many helpful discussions and Tim Tait for teaching us single tops and
carefully reading the manuscript. Johan Alwall we thank for help with
MadEvent.  Our work was supported by the Deutsche
Forschungsgemeinschaft via the Sonderforschungsbereich/Transregio
SFB/TR9 "Computational Particle Physics".


\begin{thebibliography}{99}

\bibitem{vtb}
 C.~Amsler {\it et al.}  [Particle Data Group],
  Phys.\ Lett.\  B {\bf 667}, 1 (2008);
 F.~Abe {\it et al.}  [CDF Collaboration],
  Phys.\ Rev.\ Lett.\  {\bf 74}, 2626 (1995).
 S.~Abachi {\it et al.}  [D0 Collaboration],
  Phys.\ Rev.\ Lett.\  {\bf 74}, 2632 (1995).
 V.~M.~Abazov {\it et al.}  [D0 Collaboration],
  Phys.\ Rev.\  D {\bf 78}, 012005 (2008).

\bibitem{mssm}
 H.~E.~Haber and G.~L.~Kane,
  Phys.\ Rept.\  {\bf 117}, 75 (1985).

\bibitem{mfv}
 G.~D'Ambrosio, G.~F.~Giudice, G.~Isidori and A.~Strumia,
  Nucl.\ Phys.\  B {\bf 645}, 155 (2002);
 B.~C.~Allanach, G.~Hiller, D.~R.~T.~Jones and P.~Slavich,
  JHEP {\bf 0904}, 088 (2009).

\bibitem{mrssm}
 G.~D.~Kribs, E.~Poppitz and N.~Weiner,
  Phys.\ Rev.\  D {\bf 78}, 055010 (2008);
 G.~D.~Kribs, A.~Martin and T.~S.~Roy,
  arXiv:0901.4105 [hep-ph].

\bibitem{sgluons}
 T.~Plehn and T.~M.~P.~Tait,
  J.\ Phys.\ G {\bf 36}, 075001 (2009).

\bibitem{flavor_orig}
 L.~J.~Hall, V.~A.~Kostelecky and S.~Raby,
  Nucl.\ Phys.\  B {\bf 267}, 415 (1986);
 J.~S.~Hagelin, S.~Kelley and T.~Tanaka,
  Nucl.\ Phys.\  B {\bf 415}, 293 (1994);
 F.~Gabbiani, E.~Gabrielli, A.~Masiero and L.~Silvestrini,
  Nucl.\ Phys.\  B {\bf 477}, 321 (1996);
 M.~Misiak, S.~Pokorski and J.~Rosiek,
  Adv.\ Ser.\ Direct.\ High Energy Phys.\  {\bf 15}, 795 (1998).
 M.~Ciuchini, E.~Franco, A.~Masiero and L.~Silvestrini,
  Phys.\ Rev.\  D {\bf 67}, 075016 (2003)
  [Erratum-ibid.\  D {\bf 68}, 079901 (2003)];
 A.~Bartl, K.~Hidaka, K.~Hohenwarter-Sodek, T.~Kernreiter, W.~Majerotto and W.~Porod,
  arXiv:0905.0132 [hep-ph].

\bibitem{Dittmaier:2007uw}
  S.~Dittmaier, G.~Hiller, T.~Plehn and M.~Spannowsky,
  Phys.\ Rev.\  D {\bf 77}, 115001 (2008).

\bibitem{top_fcnc}
 T.~M.~P.~Tait and C.~P.~Yuan,
  Phys.\ Rev.\  D {\bf 55}, 7300 (1997);
 J.~Cao, G.~Eilam, K.~i.~Hikasa and J.~M.~Yang,
  Phys.\ Rev.\  D {\bf 74}, 031701 (2006);
 J.~Guasch, W.~Hollik, S.~Penaranda and J.~Sola,
  Nucl.\ Phys.\ Proc.\ Suppl.\  {\bf 157}, 152 (2006);
 J.~J.~Cao, G.~Eilam, M.~Frank, K.~Hikasa, G.~L.~Liu, I.~Turan and J.~M.~Yang,
  Phys.\ Rev.\  D {\bf 75}, 075021 (2007).

\bibitem{flavor_triviality}
 J.~A.~Casas and S.~Dimopoulos,
  Phys.\ Lett.\  B {\bf 387}, 107 (1996).

\bibitem{rhogamma}
 K.~Abe {\it et al.},
  Phys.\ Rev.\ Lett.\  {\bf 96}, 221601 (2006);
 B.~Aubert {\it et al.}  [BABAR Collaboration],
  arXiv:hep-ex/0612017.

\bibitem{data}
 E.~Barberio {\it et al.}  [Heavy Flavor Averaging Group (HFAG)],
  arXiv:hep-ex/0603003.
  \url{http://www.slac.stanford.edu/xorg/hfag}

\bibitem{sfitter}
  R.~Lafaye, T.~Plehn, M.~Rauch and D.~Zerwas,
  Eur.\ Phys.\ J.\  C {\bf 54}, 617 (2008);
 R.~Lafaye, T.~Plehn, M.~Rauch, D.~Zerwas and M.~Duhrssen,
  arXiv:0904.3866 [hep-ph].

\bibitem{bsm_direct_top}
 E.~Malkawi and T.~M.~P.~Tait,
  Phys.\ Rev.\  D {\bf 54}, 5758 (1996);
 M.~Hosch, K.~Whisnant and B.~L.~Young,
  Phys.\ Rev.\  D {\bf 56}, 5725 (1997);
 T.~Han, M.~Hosch, K.~Whisnant, B.~L.~Young and X.~Zhang,
  Phys.\ Rev.\  D {\bf 58}, 073008 (1998).
 A.~Belyaev,
  arXiv:hep-ph/0007058.

\bibitem{sps}
  B.~C.~Allanach {\it et al.},
  Eur.\ Phys.\ J.\  C {\bf 25}, 113 (2002).

\bibitem{formcalc}
 T.~Hahn,
  Comput.\ Phys.\ Commun.\  {\bf 140}, 418 (2001);
 T.~Hahn and C.~Schappacher,
  Comput.\ Phys.\ Commun.\  {\bf 143}, 54 (2002).

\bibitem{looptools}
 T.~Hahn and M.~Perez-Victoria,
  Comput.\ Phys.\ Commun.\  {\bf 118}, 153 (1999).

\bibitem{hadcalc}
 M.~Rauch,
  arXiv:0804.2428 [hep-ph].

\bibitem{mlm}
 M.~L.~Mangano, M.~Moretti and R.~Pittau,
  Nucl.\ Phys.\  B {\bf 632}, 343 (2002).

\bibitem{madevent}
 J.~Alwall {\it et al.},
  JHEP {\bf 0709}, 028 (2007);
 J.~Alwall, P.~Artoisenet, S.~de Visscher, C.~Duhr, R.~Frederix, M.~Herquet and O.~Mattelaer,
  AIP Conf.\ Proc.\  {\bf 1078}, 84 (2009).

\bibitem{ckkw}
 S.~Catani, F.~Krauss, R.~Kuhn and B.~R.~Webber,
  JHEP {\bf 0111}, 063 (2001).

\bibitem{mlm_heavy}
 J.~Alwall, S.~de Visscher and F.~Maltoni,
  JHEP {\bf 0902}, 017 (2009);
 J.~Alwall, K.~Hiramatsu, M.~M.~Nojiri and Y.~Shimizu,
  arXiv:0905.1201 [hep-ph].

\bibitem{stop_jets}
 D.~O.~Carlson,
  arXiv:hep-ph/9508278;
 Z.~Sullivan,
  Phys.\ Rev.\  D {\bf 70}, 114012 (2004).

\bibitem{stop_review}
 for a new review see \eg 
 W.~Bernreuther,
  J.\ Phys.\ G {\bf 35}, 083001 (2008).

\bibitem{stop_t}
 S.~S.~D.~Willenbrock and D.~A.~Dicus,
  Phys.\ Rev.\  D {\bf 34}, 155 (1986);
 S.~Dawson and S.~S.~D.~Willenbrock,
  Nucl.\ Phys.\  B {\bf 284}, 449 (1987);
 D.~O.~Carlson and C.~P.~Yuan,
  Phys.\ Lett.\  B {\bf 306}, 386 (1993);
 T.~Stelzer, Z.~Sullivan and S.~Willenbrock,
  Phys.\ Rev.\  D {\bf 56}, 5919 (1997);
 J.~M.~Campbell, R.~K.~Ellis and F.~Tramontano,
  Phys.\ Rev.\  D {\bf 70}, 094012 (2004);
 Q.~H.~Cao, R.~Schwienhorst, J.~A.~Benitez, R.~Brock and C.~P.~Yuan,
  Phys.\ Rev.\  D {\bf 72}, 094027 (2005).

\bibitem{mcnlo}
 S.~Frixione, E.~Laenen, P.~Motylinski and B.~R.~Webber,
  JHEP {\bf 0603}, 092 (2006).

\bibitem{stop_s}
 B.~W.~Harris, E.~Laenen, L.~Phaf, Z.~Sullivan and S.~Weinzierl,
  Phys.\ Rev.\  D {\bf 66}, 054024 (2002);
 Q.~H.~Cao, R.~Schwienhorst and C.~P.~Yuan,
  Phys.\ Rev.\  D {\bf 71}, 054023 (2005).

\bibitem{stop_tw}
 T.~M.~P.~Tait,
  Phys.\ Rev.\  D {\bf 61}, 034001 (2000);
 S.~Zhu,
  arXiv:hep-ph/0109269;
 J.~M.~Campbell and F.~Tramontano,
  Nucl.\ Phys.\  B {\bf 726}, 109 (2005);
 S.~Frixione, E.~Laenen, P.~Motylinski, B.~R.~Webber and C.~D.~White,
  JHEP {\bf 0807}, 029 (2008).

\bibitem{stop_bsm}
 R.~D.~Peccei, S.~Peris and X.~Zhang,
  Nucl.\ Phys.\  B {\bf 349}, 305 (1991);
 T.~M.~P.~Tait and C.~P.~P.~Yuan,
  Phys.\ Rev.\  D {\bf 63}, 014018 (2001)'
 Q.~H.~Cao, J.~Wudka and C.~P.~Yuan,
  Phys.\ Lett.\  B {\bf 658}, 50 (2007);
 J.~A.~Aguilar-Saavedra,
  Nucl.\ Phys.\  B {\bf 804}, 160 (2008).

\bibitem{stop_analysis}
 A.~Heinson, A.~S.~Belyaev and E.~E.~Boos,
  Phys.\ Rev.\  D {\bf 56}, 3114 (1997);
 T.~Stelzer, Z.~Sullivan and S.~Willenbrock,
  Phys.\ Rev.\  D {\bf 58}, 094021 (1998);
 A.~S.~Belyaev, E.~E.~Boos and L.~V.~Dudko,
  Phys.\ Rev.\  D {\bf 59}, 075001 (1999).

\bibitem{bottom_densities}
 F.~Maltoni, Z.~Sullivan and S.~Willenbrock,
  Phys.\ Rev.\  D {\bf 67}, 093005 (2003);
 E.~Boos and T.~Plehn,
  Phys.\ Rev.\  D {\bf 69}, 094005 (2004).
 J.~M.~Campbell, R.~Frederix, F.~Maltoni and F.~Tramontano,
  arXiv:0903.0005 [hep-ph].

\bibitem{stop_angular}
 Z.~Sullivan,
  Phys.\ Rev.\  D {\bf 72}, 094034 (2005);
 G.~Mahlon,
  arXiv:hep-ph/0011349;
 P.~Motylinski,
  arXiv:0905.4754 [hep-ph].

\end{thebibliography}
\end{document}